\documentclass[12pt]{article}
\usepackage{epsfig}
\usepackage{graphics}
\usepackage{graphicx}
\usepackage[centerlast,footnotesize]{caption2}
\begin{document}
 
\begin{center}
{\bf \Large Some Features of the Production of Heavy-Quark-Containing Baryons
in Electron-Positron Collisions}

\vspace{1.12cm}
S. P. Baranov$^{\it a}$\footnote{e-mail: baranov@sci.lebedev.ru} and 
V. L. Slad$^{\it b}$\footnote{e-mail: vslad@theory.sinp.msu.ru}

\vspace{0.75cm}

$^{\it a}${\small Lebedev Institute of Physics, Russian Academy of Sciences, 
Leninskii pr. 53, Moscow 119991, Russia}

$^{\it b}${\small Skobeltsyn Institute of Nuclear Physics, Moscow State 
University, Moscow 119992, Russia}
   
\vspace{1.35cm} 
                          
{\bf Abstract}
\end{center}

\vspace*{-0.35cm}

{\small 
The production of various heavy-quark-containing baryons in electron-positron 
annihilation is considered. On the basis of exact formulas that we obtained 
previously within full perturbation theory, new numerical calculations of the 
respective cross sections are performed, and simple approximate expressions are
then constructed for the results of these calculations. The dependence of the 
total cross sections on the masses of constituent quarks is discussed. The 
application of the Peterson fragmentation function and a Reggeon-type 
fragmentation function to describing differential cross sections is analyzed.}

\vspace{1.02cm}

{\bf \center \large 1. Introduction }

Investigation of the mechanisms responsible for the production of hadrons 
containing heavy quarks is of interest from the theoretical point of view since
this provides the possibility for further testing QCD -- more precisely, our 
understanding of it. In this way, one tests both its perturbative aspect used 
to describe the simultaneous production of several quark pairs and 
nonperturbative models constructed on the basis of QCD for bound states. We 
recall that, even in cases that are the simplest at first glance, the results 
of calculations appear to be in an unexpected contradiction
with experimental data, as was, for example, in the
hadronic production of $J/ \psi$ particles. At the same
time, derivation of theoretical estimates for relevant
cross sections is of importance for practical purposes,
such as those associated with planning searches for
such particles and investigations of their properties.

Available calculations of the cross sections for
the production of baryons containing heavy $c$ and $b$
quarks rely, as a rule, on considering the production of
respective diquarks, this corresponding to the fourth
order of perturbation theory. A detailed review of the
results obtained in this way and an exhaustive list
of relevant references can be found in [1]. In the
sixth order of perturbation theory [${\cal O} (\alpha^{2} \alpha^{4}_{s})$] , 
the total and differential cross sections for the production
of multiply heavy baryons $\Omega_{scb}$ and $\Omega_{ccc}$ at the $Z$
pole in electron-positron collisions were calculated
in our previous studies [2, 3]. For the squares of
relevant matrix elements, we obtained exact analytic
expressions, which, as might have been expected,
are very cumbersome (they exist only in the form of
computer codes). As a result, numerical calculations
with these matrix elements would be extremely time-consuming.

By using these expressions and performing a series of new numerical 
calculations of various cross
sections for the production of baryons containing
three nonidentical quarks, we try here to impart, to
the emerging results, a broader content, simplicity,
and adaptability in the possible future application
to constructing estimates for planning experiments.
Specifically, we find, first of all, a simple approximate
dependence of the total cross section for baryon production in 
electron-positron collisions on the mass of
each constituent quark. On the basis of the concept
of fragmentation, we then approximate the differential
cross sections with the aid of the Peterson function
for various sets of quark masses.

In this study, we also analyze some aspects of the
description of the differential cross sections for the
production of $\Omega_{scb}$ and $\Omega_{ccc}$ baryons in terms of a
Reggeon-type fragmentation function.

\vspace{1.02cm}

{\bf \center \large 2. Dependence of the total cross sections for baryon 
production on consistent-quark masses}

Let us consider some properties of the production
of a $q_{1}q_{2}q_{3}$ baryon consisting of three nonidentical
quarks $q_{1}$, $q_{2}$  and $q_{3}$  and having a mass $M$, a momentum 
${\bf p}$, and an energy $E$ at the $Z$ pole in electron-positron collisions. 
We assume that the quark masses $m_{1}$, $m_{2}$ and $m_{3}$ differ markedly 
from each other; that is,
\begin{equation}
m_{1}^{2} \ll m_{2}^{2} \ll m_{3}^{2}.
\end{equation}

In QCD, the elementary process corresponding to
the production of such a baryon is
\begin{equation}
e^{+}+e^{-} \rightarrow q_{1}(p_{1}) + q_{2}(p_{2}) + q_{3}(p_{3}) 
+ \bar{q}_{1}(p_{4})+ \bar{q}_{2}(p_{5}) + \bar{q}_{3}(p_{6}),
\end{equation}
where the quark and antiquark 4-momenta are indicated in parentheses.

The formation of the baryon from three quarks is described in the well-known 
nonrelativistic approximation [4-6], whose details required for our purposes
are as follows. First, the velocities of the quarks forming the baryon are 
assumed to be identical. Second, the differential cross section for baryon 
production is obtained from the standard differential cross section
for the process in (2) by replacing the phase space of three quarks 
by an expression proportional to the baryon phase space; that is,
\begin{equation}
\frac{d^{3}p_{1}}{(2 \pi )^{3} \cdot 2 E_{1}}
\frac{d^{3}p_{2}}{(2 \pi )^{3} \cdot  2 E_{2}}
\frac{d^{3}p_{3}}{(2 \pi )^{3}  \cdot 2 E_{3}}
\rightarrow
\frac{|\psi (0)|^{2}}{M^{2}}
\frac{d^{3}p}{(2 \pi )^{3} \cdot  2 E},
\end{equation}
where  $\psi (0)$  is the value of the wave function at the point where 
relative coordinates of all three quarks are zero.

In each of the Feynman diagrams corresponding
to the process in (2), one can indicate a virtual gluon
$g$ such that it transforms into a quark-antiquark pair
$q_{i}(p_{i}) \bar{q}_{i}(p_{i+3})$ without emitting a new gluon $g'$. The
denominator of the propagator of the gluon $g$ -- it has
the form $(p_{i}+p_{i+3})^{2}$  -- attains a minimum of $4 m_{i}^{2}$  at 
$p_{i+3}=p_{i}$ . But if a virtual gluon $g$ transforms into two
quark-antiquark pairs 
$q_{i}(p_{i})q_{j}(p_{j})\bar{q}_{i}(p_{i+3})\bar{q}_{j}(p_{j+3})$,
the denominator of its propagator has a minimum
of  $4 (m_{i}+m_{j})^{2}$. Taking into account the inequalities
in (1), we can deduce from the above that the leading
contribution to the amplitude of the process in (2)
comes from the diagrams where the production of
quark-antiquark pairs proceeds hierarchically from
the heaviest to the lightest. This sequence that ends
up in the formation of a baryon from the quarks $q_{1}$, $q_{2}$
and  $q_3$ is generally referred to as the fragmentation of
the quark $q_3$ into a $q_{1}q_{2}q_{3}$ baryon. It is often described
analytically in the form
\begin{equation}
\frac{d\sigma}{dz} = \sigma_{q_{3}\bar{q}_{3}} \cdot D(z),
\end{equation}
where $\sigma_{q_{3}\bar{q}_{3}}$ is the total cross section for the process
$e^{+}e^{-} \rightarrow {q_{3}\bar{q}_{3}}$ and $D(z)$ is the respective 
fragmentation function. For the variable $z$, one usually takes the
quantity  $x_{p} = p/p_{\rm max}$   or   $x_{E} = E/E_{\rm max}$.

It is obvious that the amplitude of any process is
a homogeneous function of the 4-momenta $P_j$, the
masses $M_j$ , and the widths $\Gamma_j$ of real and virtual
particles involved in the process. Therefore, the total
cross section or one differential cross section or another can be represented 
in the form of the product of
some power of the total energy $\sqrt{s}$ and a function of
the reduced 4-momenta $P_j / \sqrt{s}$, the reduced masses
$M_j / \sqrt{s}$, and the reduced widths $\Gamma_j / s$. From here, it
follows, among other things, that the fragmentation
function $D(z)$ appearing in (4) depends parametrically on the reduced masses 
$m_{i} / \sqrt{s}$ of the product
quarks. In the following, we will write the reduced
masses explicitly only in the logarithmic factors on
the right-hand side of formula (6) (see below).

In comparing experimental results obtained in
electron-positron collisions with some fragmentation
function, attention is given primarily to its form
depending on one or two parameters but not to its
normalization.

For want of experimentally observed events involving the production of 
$q_{1} q_{2} q_{3}$ baryons in electron-positron annihilation, it is reasonable
to focus on the total cross section--namely, on the dependence of the
total cross section on the masses of the quarks $q_{1}$, $q_{2}$
and $q_{3}$ . The choice of a simple algebraic expression 
representing this dependence is based on the fact that
the square of the matrix element of the process being
considered is similar, in some respects, to a rational function of the momenta 
of product particles, its denominator at the minimum involving constituent
quark masses and their sums as factors. It is well known that the 
integral of such a function can generally include logarithmic terms.

By using relation (3) and the results of numerical
calculations of the total cross section for the production of 
$q_{1}q_{2}q_{3}$ baryons at the $Z$ pole in electron-positron 
collisions for six sets of masses $m_{1}, m_{2}$
and $m_{3}$ for the same set of electroweak-interaction
coupling constants (such as that for the production
of $scb$  baryons), we arrive at the formulas
\begin{equation}
\sigma_{\rm tot}=
\frac{|\psi (0)|^{2}}{(m_{1}+m_{2}+m_{3})^{2}} G,
\end{equation}
\begin{equation}
G \approx \frac{C}{m_{1}^{2}m_{2}^{2}} 
\ln \left( \frac{\sqrt{s}}{4 m_{1}} \right)
\ln\left( \frac{\sqrt {s}}{m_{3}} \right) ,
\end{equation}     
where $\sqrt{s}= 91.2$ GeV and $C=(0.0407 \pm 0.0006)$ pb.

In Table 1, we present the sets of masses $m_{1}$, $m_{2}$
and  $m_{3}$; the values of $G$ that are obtained from a
Monte Carlo calculation of the integral of the square
of the matrix element for the process in (2) over the
phase space of four final particles (QCD column in
the table); and the values of $G$ that are obtained by
formula (6).

We believe that the use of the factor $ln( \sqrt{s}/(4m_{1} ))$
in formula (6) is quite justified empirically. It seems
plausible that the quantity $G$ depends only slightly
on the mass $m_{3}$ of the heaviest quark; relying on
the results of our calculations exclusively, we cannot be confident, 
however, that the $G$ depends on $m_3$ through the 
factor $ln( \sqrt{s}/m_{3} )$, as follows from Eq. (6).

{\bf Table 1}
{\normalsize
\begin{center} 
\begin{tabular}{|p{1.15cm}| p{1.15cm}| p{1.15cm}| p{3.45cm}| p{3.45 cm}| 
p{2.5 cm}|} \hline
$m_{1}$ & $m_{2}$ & $m_{3}$ & QCD, $G$ & Formula (6), $G$ & Parametr $\epsilon$
 \\
(GeV) & (GeV) & (GeV) & (pb $\cdot$ GeV$^{4}$) & (pb $\cdot$ GeV$^{4}$)
&from  (8) \\ \hline
0.5 & 1.5 & 4.8  &$(7.85 \pm 0.16) \cdot 10^{-1}$ 
& $(8.15 \pm 0.14)\cdot 10^{-1}$ & $0.124 \pm 0.016$ \\  
0.3 & 1.5 & 4.8 & $(2.58 \pm 0.07) \cdot 10^{0}$ 
& $(2.57 \pm 0.04) \cdot 10^{0}$ & $0.098 \pm 0.012$ \\
0.075 & 1.5 & 4.8 & $(5.59 \pm 0.20 ) \cdot 10^{1}$
& $(5.42 \pm 0.09 ) \cdot 10^{1}$ & $0.066 \pm 0.008$ \\
0.01 & 1.5 & 4.8 & $(4.29 \pm 0.12 ) \cdot 10^{3}$
& $(4.12 \pm 0.07 )\cdot 10^{3}$ & $0.048 \pm 0.006$ \\
0.01 & 0.5 & 4.8 & $(3.90 \pm 0.28 ) \cdot 10^{4}$
& $(3.71 \pm 0.06 )\cdot 10^{4}$ & $0.016 \pm 0.002$ \\ 
0.01 & 0.5 & 1.5 & $(5.10 \pm 0.53 ) \cdot 10^{4}$
& $(5.18 \pm 0.09 )\cdot 10^{4}$ & $0.24 \pm 0.08$ \\ \hline
\end{tabular}
\end{center}
}

Strictly speaking, the dependence of the total
cross section on the quark masses $m_{1}$, $m_{2}$ and $m_{3}$
is not exhausted by the explicit expressions in
formulas (5) and (6), since the baryon-state wave
function at the origin,  $\psi (0)$, must change in response
to a change in the masses. The dependence of $\psi (0)$ on
the masses of constituent quarks is determined within
potential quark models, which are not discussed in
the present study. On the basis of the numerical
values of $|\psi (0)|^{2}$ that are presented in [7] for six sets
of quarks $q_{1}$, $q_{2}$, and $q_{3}$, we can nevertheless estimate
cross sections by using the approximate expression
\begin{equation}
|\psi (0)|^{2} \approx D m_{2}^{1.5} m_{3}, \hspace{0.5cm} {\mbox{\rm if}}
\hspace{0.3cm} if \hspace{0.2cm} m_{1} \ll m_{2} \leq m_{3},
\end{equation}
where $D = 0.065 \cdot 10^{-3}$ GeV$^{3.5}$.

Taken together, relations (5)-(7) indicate that
the total cross section for the production of $q_{1}q_{2}q_{3}$
baryons in electron-positron annihilation is highly
sensitive to the mass of the lightest of three quarks.
This circumstance is especially important in the case
where, for the quark $q_{1}$, one takes a  $u$  or a $d$ quark,
since, from the point of view of simple nonrelativistic
concepts, their masses can be varied within rather
wide intervals, from 50 MeV (in pions) to 300 MeV (in
nucleons). In order to estimate cross sections for the
production of baryons containing two heavy quarks
and a  $u$  or a $d$ quark, we set $m_{u}=m_{d}= 100$  MeV
in Eqs. (5)-(7), bearing in mind that, according to
the approximation specified by Eq. (6), these cross
section are determined to within a factor of 10.

For want of something better and, to some extent,
as a continuation of the strategies adopted in [7], we
propose extending the procedure of the present study
and of previous studies [2, 3] to the case of deriving
estimates for the production of baryons containing
one heavy quark $c$ or $b$ , a strange quark $s$, and a light
quark $u$ or $d$--namely, we propose supplementing
the perturbative part of calculations (sixth order of
perturbation theory) with the following nonrelativistic
nonperturbative elements: the assumption of equal
velocities of the quarks fusing into the baryon in question, 
relation (3), and the approximation specified by
Eq. (7). What we inherit from nonrelativistic potential
models reduces to the extrapolations in (3) and (7).
It is reasonable to indicate here that, in contrast to $u$
and $d$ quarks, a strange quark of mass $m_{s}=500$  MeV
leads to an acceptable level of agreement with naive
nonrelativistic expectations for the masses of the meson 
and baryon ground states [$s\bar{s}$ ($\phi$) and $sss$ ($\Omega$),
respectively].
 
All estimates that can be obtained on the basis
of formulas (5)-(7) by setting $m_{b}=4.8$ GeV, $m_{s}=0.5$ GeV, 
and  $m_{u}=m_{d}=0.1$ GeV are given in Table 2 (the symbol $q$ 
in the subscripts there stands for a
$u$ or a $d$ quark). The factor $C$ in (6) is set to 0.0407 pb
in calculating cross sections for the production of
$\Omega_{scb}$ , $\Xi_{qcb}$ , and $\Xi_{qsb}$ baryons and to $C$ = 0.0407 pb·
$ \cdot ((g_{V}^{c})^{2}+g_{A}^{c})^{2})/
((g_{V}^{b})^{2}+g_{A}^{b})^{2}$ = 0.0317 pb in dealing 
with $\Xi_{qsc}$ baryons.

\hspace{1.0cm} {\bf Table 2} 
{\normalsize
\begin{center}
\begin{tabular}{|p{4.0cm}| p{4.0cm}| p{4.0cm}| } \hline
Baryon & $10^{3} \cdot | \psi (0)|^{2}$, GeV$^{6}$  & $\sigma_{\rm tot}$, fb \\
\hline 
$\Omega_{scb}$   & 0.57    & 0.0097  \\  
$\Xi_{qcb}$   & 0.57    & 0.40  \\
$\Xi_{qsb}$   & 0.11    & 0.98  \\
$\Xi_{qsc}$   & 0.034   & 2.2   \\  \hline
\end{tabular}
\end{center}
}

We would like to bring to the attention of the reader
that, in [8], the quantity $|\psi (0)|^{2}$ was calculated at
the following values of the constituent quark masses:
$m_b$ = 5.29 GeV, $m_c$ = 1.905 GeV, $m_s$ = 0.6 GeV, and
$m_u$ = $m_d$ = 0.3 GeV. In our previous studies [2, 3],
we employed the values of $|\psi (0)|^{2}$ from [8] without
introducing any corrections. In the present study, all
cross-section values, including those that are given in
the figures, are calculated by using Eq. (7).

It should be emphasized that the approximate
formula (6) was obtained for the constituent quark
masses obeying the inequalities in (1). But if these
inequalities are not satisfied and if, in addition, there
are quarks identical in flavor among the quarks in
question, formula (6) cannot be used even for a rough
estimate of cross sections. Indeed, it can be seen that,
for the production of an $\Omega_{ccc}$ baryon in 
electron-positron annihilation, a direct numerical calculation
of the quantity $G$ on the basis of formula (5) gives
the value of 2.27 pb GeV$^4$, while expression (6)
yields 0.090 pb GeV$^4$. The reasons for so significant
a discrepancy between the values of $G$ are rather
obvious: if the inequalities in (1) hold, the main
contribution to the cross section comes only from a
few Feynman diagrams, but, if all three masses $m_1$,
$m_2$, and $m_3$ are close to one another, all diagrams
make approximately equal contributions (for the
production of an $\Omega_{ccc}$ baryon, there are 504 such
diagrams); additionally, interference effects arise if
there are identical quarks.

\vspace{1.02cm}

{\bf \center \large  3. Differential cross section
and Peterson fragmentation function }

Let us now proceed to consider a simple algebraic description of the 
differential cross sections for
the production of $q_{1}q_{2}q_{3}$  baryons in electron-positron
collisions on the basis of the fragmentation approach.
Experimental data on the production of heavy-quark
hadrons in electron­positron annihilation are usually
approximated in terms of the Peterson function [9]
\begin{equation}
D(z) \sim \frac{1}{z[1-(1/z) - \varepsilon/(1-z)]^{2}}
\end{equation}
where $\varepsilon$ is a phenomenological parameter. We used
this function in [2] to describe approximately the production 
of $\Omega_{scb}$ baryons in electron-positron collisions.

We performed complete numerical calculations
of the differential cross sections for the production
of $q_{1}q_{2}q_{3}$ baryons for six sets of constituent quark
masses (see Table 1) and then determined the values
of the parameter $\varepsilon$ of the Peterson function (8) that
ensure the best agreement between the form of this
fragmentation function and the form of the differential
cross sections $d\sigma / dx_{p}$. The resulting values of the
parameter  are given in the last column of Table 1.
The results of numerical calculations of $d\sigma / dx_{p}$ and
their approximation in terms of the Peterson function
are shown in Fig. 1.
    
The dependence of the parameter $\varepsilon$ on the mass $m_1$
of the lightest quark can be approximated by a linear function,
\begin{equation}
\varepsilon \approx a +b m_{1}
\end{equation}
where $a = 0.046$ and $b = 0.17$ GeV$^{-1}$. As the middle
mass $m_{2}$ is decreased, the maximum of the distribution
$d\sigma / dx_{p}$ is shifted toward the largest possible 
relative momentum value, $x_{p} = 1$, while the parameter
$\varepsilon$ becomes smaller. The shift of the maximum of the
cross section $d\sigma / dx_{p}$ in response to the change in the
mass $m_{3}$ of the heaviest quark is opposite to that in
response to the analogous change in $m_{1}$ and $m_{2}$: with
decreasing $m_{3}$, the extremal value of $x_{p}$ decreases
substantially, while the parameter $\varepsilon$ in the function
given by (8) grows.

The value $\varepsilon$ that we found here for  at the mass
values of $m_{1}=0.01$ GeV, $m_{2}=0.5$ GeV, and  $m_{3}=1.5$ GeV 
is very close to the values of $\varepsilon$  that were
obtained in experiments that studied the production
of $\Lambda_{c}$, $\Sigma_{c}$, and $\Xi_{c}$
baryons in electron-positron annihilation: 
$\varepsilon=0.236^{+0.068}_{-0.048}$ for $\Lambda_{c}$ [10],
$\varepsilon= 0.29 \pm 0.06$ for $\Sigma_{c}$ [11],
$\varepsilon= 0.24 \pm 0.08$ for $\Xi_{c}$ [12].
    
For baryons containing quarks of identical flavor
(for example, $\Omega_{ccc}$ ), it is difficult to take into account
effects of interference between identical particles;
therefore, the fragmentation mechanism cannot be so
well justified theoretically in this case as for $q_{1}q_{2}q_{3}$
baryons under the conditions in (1). At the same
time, it remains quite useful in the phenomenological
aspect. For example, it was found in [3] that the
transverse-momentum distribution of $\Omega_{ccc}$ baryons
that is obtained from direct numerical calculations
cannot be adequately approximated with the aid of
the Peterson function. However, the shape of this
distribution can be faithfully reproduced by using the
so-called Lund function [13]
\begin{equation}
D(z) \sim \frac{1}{z} (1-z)^{a} \exp (-c/z),
\end{equation}
where the parameters are set to a = 2.4 and c = 0.70.

\vspace{1.02cm}

{\bf \center \large  4. Differential cross sections
and Reggeon-type fragmentation function }

Let us now consider the representation of numerical results for the 
production of heavy-quark-containing baryons in electron-positron collisions
in terms of the Reggeon-type fragmentation function [14]
\begin{equation}
D(z) \sim z^{\beta} (1-z)^{\gamma}.
\end{equation}
This function was employed in [15, 16] in discussing
experimental data on the production of $D$ and $B$
mesons in electron­positron annihilation, the 
parameters $\beta$ and $\gamma$ not being related there to 
Regge trajectories.
    
An approximation of the numerical results obtained here 
for the production of $\Omega_{scb}$ and $\Omega_{ccc}$ baryons
in electron-positron annihilation with the aid of the
fragmentation function in (11) was performed for the
differential cross sections $d\sigma / d x_{E}$ and is displayed in
Fig. 2.
    
It is of interest to compare the values obtained for
the parameters of the function in (11) by approximating the 
numerical results of our perturbative approach to 
determining cross sections and the values
deduced from the expressions for these parameters in
terms of the intercepts of the trajectories of appropriate 
hadrons. If the quark $i$ fragments into a baryon
consisting of the quarks $i, j$, and $k$, then [17]
\begin{equation}
\beta=1-\alpha_{i\bar{i}}, \hspace{1.0 cm} \gamma=
\alpha_{q\bar{q}} - 2\alpha_{jkq},
\end{equation}
where the symbol $q$ in the subscripts stands for a
light quark $u$ or $d$, while the quantities $\alpha_{l\bar{l}}$ and 
$\alpha_{jkq}$ are the intercepts of the meson (of $\alpha_{l\bar{l}}$ quark 
content) and baryon (of $\alpha_{jkq}$ quark content) trajectories. An
approximate linear relation between the intercepts
of trajectories corresponding to different quark 
contents was obtained within the model of quark-gluon strings [18]:
\begin{equation} 
2(\alpha_{ijk} - \alpha_{ijl}) = \alpha_{k\bar{k}} -  \alpha_{l\bar{l}}.
\end{equation}
      
On the basis of relation (13) and the standard notation 
$\alpha_{u\bar{u}} = \alpha_{d\bar{d}} = \alpha_{\rho}$,
$\alpha_{s\bar{s}} = \alpha_{\phi}$, $\alpha_{c\bar{c}} = \alpha_{\psi}$,
$\alpha_{b\bar{b}} = \alpha_{\Upsilon}$, and
$\alpha_{uud} = \alpha_{udd} = \alpha_{N}$
the equalities in (12) can be recast into the form
\begin{equation}
\beta  =  1-\alpha_{\Upsilon}, \hspace{1.0 cm} \gamma=
3 \alpha_{\rho} - \alpha_{\phi} - \alpha_{\psi} - 2 \alpha_{N},
\end{equation}
for $\Omega_{scb}$ baryons and into the form
\begin{equation}
\beta  =  1-\alpha_{\psi}, \hspace{1.0 cm} \gamma=
3 \alpha_{\rho} - 2 \alpha_{\psi}  - 2 \alpha_{N},
\end{equation}
for $\Omega_{ccc}$ baryons.

For some intercepts, we will take the quite reliably
established values of $\alpha_{N} = -0.4$, $\alpha_{\rho} = 0.5$, and  
$\alpha_{\phi} \approx 0$ [17], while, for the others, we will use the 
following estimates: $\alpha_{\psi} = -2.2$ [19] and 
$\alpha_{\Upsilon} = -8.0$ [20]. These
estimates differ only slightly from those that were
previously obtained in [7, 21].

Substituting the above values into relation (14)
for $\Omega_{scb}$ baryons, we obtain $\beta = 9.0$ $\gamma = 4.5$, but
these results are in a glaring contradiction with the
values of  $\beta = 3.3$ and $\gamma = 1.48$ if  $m_{s}=500$ MeV
and with the values of  $\beta = 3.2$ and $\gamma = 1.18$ if 
$m_{s}=300$, which we obtained from a comparison of
the fragmentation function in (11) with the results
of direct numerical calculations of the cross section
$d\sigma / dx_{E}$. In turn, relation (15) for the production of
$\Omega_{ccc}$ baryons yields  $\beta = 3.2$ and $\gamma = 6.7$, 
which can be thought to be in very rough agreement with the values
$\beta = 2.6$ and $\gamma = 4.6$ resulting from the 
approximation of our numerical results for the cross section
$d\sigma / dx_{E}$.

This suggests the simple conclusion that the
perturbative and the nonperturbative (Reg- geon) 
approach do not reduce to each other and, depending on
the process being considered, they can lead either to
close or to strongly different results.

\newpage

\begin{figure}[ht]
\vspace*{-2.95cm}  \hspace*{-1.2cm}
\includegraphics[scale=0.835]{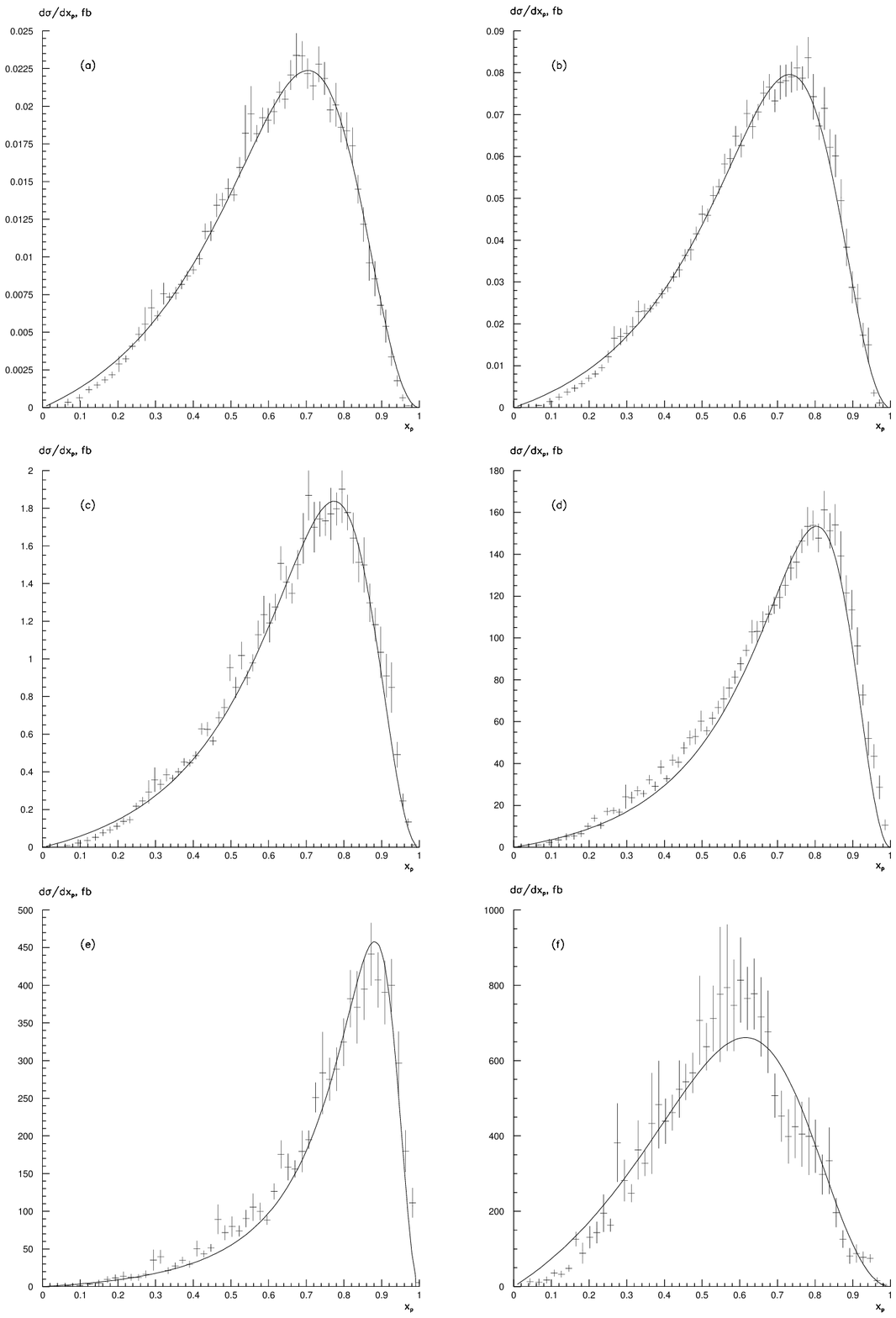}
\vspace*{-5.3cm}\\
{\footnotesize{\bf Fig. 1.} {Differential cross section  $d \sigma / d x_{p}$
for the production of $q_{1}q_{2}q_{3}$ baryons  in electron-positron 
collisions:  (crosses) results of Monte Carlo calculations along with the 
errors in them and (solid curves) results of the calculation by formula (4)
with the Peterson fragmentation function (8). The values of the masses of the 
quarks $q_{1}$, $q_{2}$, and $q_{3}$ (in GeV) and of the parameter $\varepsilon$ 
in the Peterson function in Figs. 1{\it a} -- 1{\it f} \ are the following: 
({\it a}) $m_{1} = 0.5$ GeV, $m_{2} = 1.5$ GeV, $m_{3} = 4.8$ 
GeV, $\varepsilon = 0.124$; ({\it b}) $m_{1} = 0.3$ GeV, $m_{2} = 1.5$ GeV, 
$m_{3} = 4.8$ GeV, $\varepsilon = 0.098$; ({\it c}) $m_{1} = 0.075$ GeV, 
$m_{2} = 1.5$ GeV, $m_{3} = 4.8$ GeV, $\varepsilon = 0.066$;
({\it d}) $m_{1} = 0.01$ GeV, $m_{2} = 1.5$ GeV, $m_{3} = 4.8$ GeV,
$\varepsilon = 0.048$; ({\it e}) $m_{1} = 0.01$ GeV, $m_{2} = 0.5$ GeV, 
$m_{3} = 4.8$ GeV, $\varepsilon = 0.016$; ({\it f}) $m_{1} = 0.01$ GeV, 
$m_{2} = 0.5$ GeV, $m_{3} = 1.5$ GeV, $\varepsilon = 0.24$.}}
\end{figure}

\newpage

\begin{figure}[ht]
\vspace*{-9.7cm}  \hspace*{-1.2cm}
\includegraphics[scale=0.835]{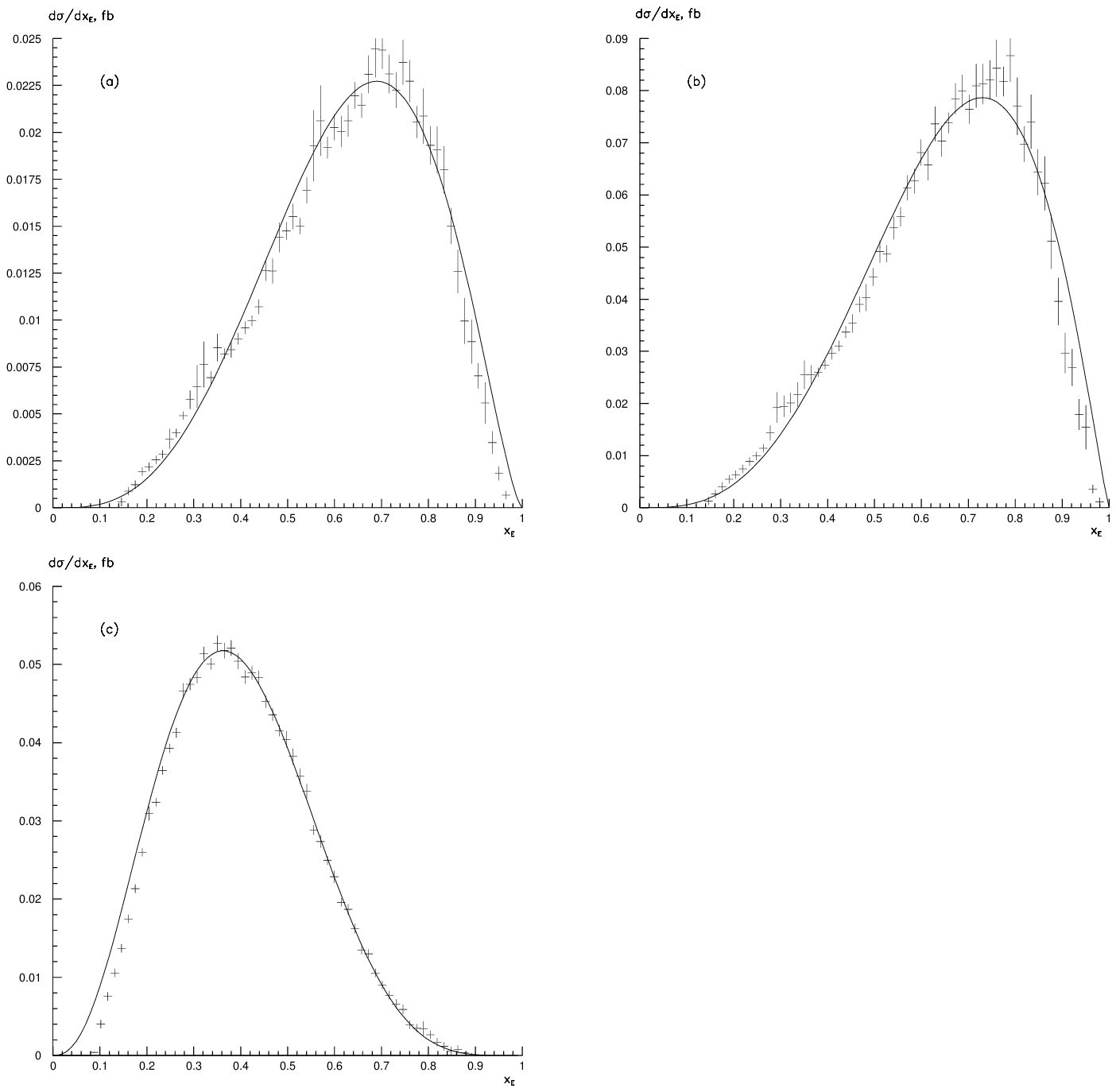}
\vspace*{-11.25cm}\\
{\footnotesize{\bf Fig. 2} {Differential cross section $d \sigma / d x_{E}$ for
the production of $\Omega_{scb}$ baryons [({\it a}) $m_{s} = 0.5$ GeV and 
({\it b}) $m_{s} = 0.3$ GeV] and $\Omega_{ccc}$ baryons ({\it c}) in 
electron-positron annihilation. The crosses show the results of Monte Carlo 
calculations and their errors. The solid curves correspond to the approximation
of the cross sections with the aid of a Reggeon-type fragmentation function 
(11) with the parameters ({\it a}) $\beta = 3.3$ and $\gamma = 1.48$, ({\it b})
$\beta = 3.2$ and $\gamma = 1.18$, and ({\it c}) $\beta = 2.6$ and 
$\gamma = 4.6$.}}
\end{figure}


\begin{thebibliography}{99}

\bibitem{1}
 V. V. Kiselev and A. K. Likhoded, {\bf Usp. Fiz. Nauk 172},
   497 (2002) [Phys. Usp. 45, 455 (2002)].

\bibitem{2}
 S. P. Baranov and V. L. Slad, {\bf Yad. Fiz. 66}, 1778 (2003)
    [Phys. At. Nucl. 66, 1730 (2003)].

\bibitem{3}
 S. P. Baranov and V. L. Slad, {\bf Yad. Fiz. 67}, 829 (2004)
    [Phys. At. Nucl. 67, 808 (2004)].

\bibitem{4}
 C.-H. Chang, {\bf Nucl. Phys. B 172}, 425 (1980).

\bibitem{5}
 R. Baier and R. Ruckl, {\bf Phys. Lett. B 102}, 384 (1981).

\bibitem{6}
D. Jones, {\bf Phys. Rev. D 23}, 1521 (1981).

\bibitem{7}
V. G. Kartvelishvili and A. K. Likhoded, {\bf Yad. Fiz. 29},
    757 (1979) [Sov. J. Nucl. Phys. 29, 390 (1979)].

\bibitem{8}
E. Bagan, H. G. Dosch, P. Godzinsky, et al., {\bf Z. Phys.
     64}, 57 (1994).

\bibitem{9}
C. Peterson, D. Schlatter, J. Schmitt, and P. M. Zer-
    was, {\bf Phys. Rev. D 27}, 105 (1983).

\bibitem{10}
ARGUS Collab. (H. Albrecht et al.), {\bf Phys. Lett. B
    207}, 109 (1988).

\bibitem{11}
ARGUS Collab. (H. Albrecht et al.), {\bf Phys. Lett. B
    211}, 489 (1988).

\bibitem{12}
 ARGUS Collab. (H. Albrecht et al.), {\bf Phys. Lett. B
    247}, 121 (1990).

\bibitem{13}
B. Andersson, G. Gustafson, and B. Soderberg,
    {\bf Z. Phys. C 20}, 317 (1983).

\bibitem{14}
 V. G. Kartvelischvili, A. K. Likhoded, and V. A. Petrov,
    {\bf Phys. Lett. B 78}, 615 (1978).

\bibitem{15}
 G. Colangelo and P. Nason, {\bf Phys. Lett. B 285}, 167
    (1992).

\bibitem{16}
 OPAL Collab. (G. Alexander et al.), {\bf Phys. Lett. B
    364}, 93 (1995).

\bibitem{17}
 A. B. Kaidalov, {\bf Yad. Fiz. 45}, 1452 (1987) [Phys. At.
    Nucl. 45, 902 (1987)].

\bibitem{18}
 A. B. Kaidalov, {\bf Z. Phys. C 12}, 63 (1982).

\bibitem{19}
 A. B. Kaidalov and O. I. Piskunova, {\bf Yad. Fiz. 43}, 1545
    (1986) [Sov. J. Nucl. Phys. 43, 994 (1986)].

\bibitem{20}
 O. I. Piskunova, {\bf Yad. Fiz. 57}, 538 (1994) [Phys. At.
    Nucl. 57, 508 (1994)].

\bibitem{21}
 V. G. Kartvelischvili and A. K. Likhoded, {\bf Yad. Fiz. 42},
    1306 (1985) [Sov. J. Nucl. Phys. 42, 823 (1985)].


\end{thebibliography}
\end{document}